\documentclass[%
preprint,
superscriptaddress,
 amsmath,amssymb,
 aps,
 showkeys,
 titlepage
]{revtex4-2}

\usepackage{graphicx}
\usepackage{dcolumn}
\usepackage{bm}
\usepackage[mathlines]{lineno}

\usepackage[colorlinks = true,
            linkcolor = blue,
            urlcolor  = blue,
            citecolor = red,
            anchorcolor = blue]{hyperref}

\begin{document}


\title{Significant Phonon Drag Effect in Wide Bandgap GaN and AlN} 

\author{Yujie Quan}
\affiliation{Department of Mechanical Engineering, University of California, Santa Barbara, CA 93106, USA}
\author{Yubi Chen}
\affiliation{Department of Mechanical Engineering, University of California, Santa Barbara, CA 93106, USA}
\affiliation{Department of Physics, University of California, Santa Barbara, CA 93106, USA}

\author{Bolin Liao}
\email{bliao@ucsb.edu} \affiliation{Department of Mechanical Engineering, University of California, Santa Barbara, CA 93106, USA}

\date{\today}

\begin{abstract}
A thorough understanding of electrical and thermal transport properties of group-III nitride semiconductors is essential for their electronic and thermoelectric applications. Despite extensive previous studies, these transport properties were typically calculated without considering the nonequilibrium coupling effect between electrons and phonons, which can be particularly strong in group-III nitride semiconductors due to the high electric fields and high heat currents in devices based on them. In this work, we systematically examine the phonon drag effect, namely the momentum exchange between nonequilibrium phonons and electrons, and its impact on charge mobility and Seebeck coefficient in GaN and AlN by solving the fully coupled electron and phonon Boltzmann transport equations with \textit{ab initio} scattering parameters. We find that, even at room temperature, the phonon drag effect can significantly enhance mobility and Seebeck coefficient in GaN and AlN, especially at higher carrier concentrations. Furthermore, we show that the phonon drag contribution to mobility and Seebeck coefficient scale differently with the carrier concentration and we highlight a surprisingly important contribution to the mobility enhancement from the polar optical phonons. We attribute both findings to the distinct mechanisms the phonon drag affects mobility and Seebeck coefficient. Our study advances the understanding of the strong phonon drag effect on carrier transport in wide bandgap GaN and AlN and gives new insights into the nature of coupled electron-phonon transport in polar semiconductors.

\end{abstract}

\keywords{Phonon Drag, Electron-phonon Coupling, Transport Coefficient}
                            
\maketitle



\section{Introduction}
Recently, the development of group-III nitride semiconductors, including GaN, AlN, and their alloys, has made a significant impact on a broad range of applications such as solar cells~\cite{matioli2011highsc2}, light-emitting diodes~\cite{khan2008ultravioletLED4,wasisto2019beyondLED1,denbaars2013developmentLED2,susilo2018alganLED3}, and photodetectors~\cite{gundimeda2017fabricationPD,chowdhury2019selfPD2}. Moreover, the high carrier mobility, which is as high as $\rm 1200~cm^2/V\,s$ in GaN~\cite{look2001predictedGaNmax} and $\rm 400~cm^2/V\,s$ in AlN\cite{taniyasu2006increasedAlN426} at room temperature, together with high breakdown electric fields and high thermal conductivities make these two materials excellent candidates for high-power and high-frequency electronic devices~\cite{wong2013nhighfreq,chowdhury2013lateral}. However, the self-heating effect due to the involved high power density is one of the main limitations of the performance of the III-nitride semiconductors~\cite{hassan2018ganheat,fletcher2017surveyheat2}. Thanks to their high Seebeck coefficient, it is promising to integrate on-chip thermoelectric spot cooling~\cite{chen2022thermoelectriccooling} in the proximity of a high-power transistor with the same III-nitride material, which provides a solution for heat dissipation without adding new materials. In addition, their intrinsically high Seebeck coefficient, large band gap and high temperature stability of their electrical properties make them good candidates for thermoelectric power generation devices at elevated temperatures~\cite{lu2013iii,sztein2013calculatedAlNBTE} Therefore, a thorough theoretical understanding of the coupled electrical, thermal and thermoelectric transport in these materials is critical for the electrical-thermal codesign of devices based on them~\cite{choi2021perspective}. 

There have been extensive theoretical studies of electrical, thermal and thermoelectric transport properties of the group-III nitrides. Like other semiconductors, their intrinsic electrical transport properties are limited by electron-phonon interactions~\cite{ziman2001electrons}. In addition to the short-ranged deformation potential mechanism, the strong polar nature of the group-III nitrides features significant polar optical phonon scattering, mediated by the long-ranged Fr\"{o}hlich dipole interaction~\cite{frohlich1954electronsFRO}. Furthermore, the wurtzite structure of the group-III nitrides lacking the inversion symmetry also gives rise to important piezoelectric scattering by the acoustic phonon modes~\cite{zook1964piezoelectric}. Early mobility calculations were based on Monte Carlo simulation, which took polar optical phonon, piezoelectric, deformation potential, and ionized impurity scatterings into account\cite{farahmand2001monte,foutz1999transient}. Analytical models considering various electron scattering mechanisms based on the electron Boltzmann transport equation (BTE) with the relaxation time approximation (RTA) were also developed\cite{abdel2005modeling,liu2005thermoelectric,sztein2013calculatedAlNBTE}. However, these methods often relied on empirical and experimental parameters and the physical insights into the transport details provided by these studies were usually limited. Recent advancements in \textit{ab initio} methods based on the density functional theory (DFT) have made it possible to directly evaluate the electron-phonon scattering rates associated with each scattering channel from first principles~\cite{giustino2017electron}. Jhalani et al. calculated the electron-phonon scattering rates of electrons and holes in GaN from first principles and simulated the cooling process of hot electrons and holes~\cite{jhalani2017ultrafast}. In a follow-up work, they also showed the importance of the dynamic quadrupolar interaction on the piezoelectric electron-phonon scattering in GaN~\cite{jhalani2020piezoelectricQD}. In parallel, Ponc\'{e} et al. calculated the electron and hole mobilities in GaN limited by electron-phonon interactions from first principles and predicted that the hole mobility can be improved by strain~\cite{ponce2019route,ponce2019hole}. On the thermal transport side, modeling thermal transport by solving the phonon BTE with interatomic force constants evaluated from DFT and density functional perturbation theory (DFPT) is now routine, and has been applied to understand the thermal conductivity of GaN~\cite{lindsay2012thermal,garg2018spectral} with good agreement with experiments. Utilizing first-principles phonon calculations coupled with the modern theory of polarization, we recently showed that thermal transport in GaN can be modulated by strong external electric fields~\cite{quan2021electric}. Despite the remarkable progress, one limitation of existing first-principles calculations of electrical and thermal transport in group III-nitrides is that the interactions between the nonequilibrium populations of electrons and phonons are ignored. Namely, phonons are assumed to be in thermal equilibrium when the scattering rates of electrons are calculated and vice versa. Due to the existence of high electric fields and high heat currents in group-III nitride-based devices, which tends to result in highly nonequilibrium electron and phonon distributions, the effect of nonequilibrium phonons on electronic transport properties, known as the phonon drag\cite{gurevich1989electron,herring1954theory}, is critical to evaluating the performance of these devices.     



Phonon drag refers to the momentum exchange between non-equilibrium phonons and electrons and is typically suppressed at higher temperatures due to the more predominant anharmonic phonon-phonon scattering than electron-phonon scattering, which hinders the momentum flow between electrons and phonons~\cite{herring1954theory}. The phonon drag contribution to the Seebeck coefficient was first recognized in germanium~\cite{frederikse1953thermoelectric,geballe1954seebeckGe}, and then in silicon~\cite{geballe1955seebeckSi} and $\rm ~FeSb_2$\cite{bentien2007colossalFeS2,pokharel2013phononFeS2,battiato2015unifiedFeS2}. In all these cases, the phonon drag contribution is prominent only at very low temperatures. The nonnegligible phonon drag contribution to the Seebeck coefficient at room temperature was first recognized in silicon by Mahan et al\cite{mahan2014seebeck}, who combined first-principles phonon calculations with an analytical electron-phonon interaction model. Zhou et al. calculated the Seebeck coefficient in silicon including the phonon drag by first-principles calculations and found that the phonon drag contributes to more than 30\% of the total Seebeck coefficient at room temperature~\cite{zhou2015ab}.  Both of these calculations solved the partially decoupled electron BTE with the assumption that the non-equilibrium distribution of the electronic system does not affect the phonon system. A similar approach was used by Bonini et al. to evaluate thermoelectric transport properties in silicon and diamond~\cite{fiorentini2016thermoelectric,macheda2018magnetotransport}. Recently, Protik et al. developed a coupled electron-phonon BTE solver framework~\cite{protik2020coupledGaAs,protik2022elphbolt}, where the mutual drag between electrons and phonons is fully captured. Using this framework, they found that in n-doped 3C-SiC, the phonon drag contributes to more than 50\% of the total Seebeck coefficient even at room temperature ~\cite{protik2020electronSiC}. Using the same method, Li et al. found that in p-doped diamond, the Seebeck coefficient is enhanced by more than a factor of 2 at 300\,K when the phonon drag is included ~\cite{li2022colossal}. In a related work, they identified an unusually large phonon drag contribution to the Seebeck coefficient one order of magnitude higher than the normal diffusive Seebeck coefficient in heavily doped p-type cubic boron arsenide~\cite{li2023high}. Experimentally, a recent study in AlGaN/GaN two-dimensional electron gas showed that the phonon drag contributes to 32\% of the Seebeck coefficient at room temperature~\cite{yalamarthy2019significant2D}. In addition to the Seebeck coefficient, the increase of carrier mobility at 300\,K due to the phonon drag effect was also predicted in 3C-SiC ~\cite{protik2020electronSiC} and GaAs~\cite{protik2020coupledGaAs}. 

In this work, we utilized the computational framework developed by Protik et al.~\cite{protik2022elphbolt} to investigate the influence of the phonon drag effect on electrical transport properties in both n-type and p-type GaN and AlN. We focused our analysis on n-type transport with electron concentrations ranging from $\rm 10^{15}~cm^{-3}$ to $\rm 10^{19}~cm^{-3}$. We found that, at room temperature, the phonon drag effect has little contribution to the electron mobility in GaN and AlN at low doping levels, while its contribution becomes more evident with the increasing carrier concentration. Besides, a significant enhancement of the Seebeck coefficient due to phonon drag was found in both GaN and AlN throughout the carrier concentration range that we investigated. The microscopic mechanisms of the phonon drag contribution to the mobility and the Seebeck coefficient were also analyzed. Our work provides a detailed fundamental understanding of the phonon drag effect and its impact on electrical transport properties in wide bandgap group-III nitrides. 

\section{Computational Methods}

\subsection{Density Functional Theory Calculations}
First principles electronic structure calculations were carried out using the Quantum ESPRESSO (QE) package~\cite{giannozzi2009quantumQE} with the scalar-relativistic Optimized Norm-Conserving Vanderbilt (ONCV) pseudopotentials~\cite{hamann2013optimizedONCV} within the local density approximation (LDA)~\cite{perdew1992accurateLDA}. The kinetic energy cutoff for wavefunctions was set to $\rm 80~Ry$. A mesh grid of $12\times 12\times 12$ in the first Brillouin zone (BZ) was adopted and the total electron energy convergence threshold for self-consistency was set to $\rm 1 \times10^{-10}~Ry$. The crystal lattice was fully relaxed with a force threshold of $\rm 10^{-4} ~eV/\textup{\AA}$, with lattice parameters $a = 3.16~\textup{\AA} $, $c = 5.148~\textup{\AA} $ in GaN and $a= 3.12 ~\textup{\AA}$, $c = 5.0~\textup{\AA}$ in AlN, both of which were in excellent agreement with the experimental values~\cite{schulz1977crystal}. The phonon dispersion was calculated using DFPT~\cite{baroni2001phononsDFPT} with a threshold of $\rm 10^{-13}~Ry$ for self-consistency on a $6\times6\times6$ q-point grid. The non-analytical correction term due to the long-range Coulomb interactions (the Fr\"{o}lich interaction) was also included. The third-order anharmonic interatomic force constants were computed using a $3\times3\times3$ supercell using the finite displacement method~\cite{li2014shengbte}, taking up to the fifth nearest neighbors into consideration. The electron-phonon coupling were calculated using the EPW code~\cite{ponce2016epw}, where the electron-phonon matrix elements were first calculated on a coarse $12\times 12 \times12$ k-point grid and $6\times6\times6$ q-point grid and then transformed to the real-space Wannier representation.

\subsection{Phonon Drag and Coupled Electron-Phonon BTEs}
To capture the phonon drag contribution to electrical transport properties, coupled electron-phonon BTEs with electron-phonon matrix elements calculated from first principles need to be solved. The steady-state electron and phonon BTEs can be written as~\cite{zhou2016first}:
\begin{equation} \label{eqn:1}
\left\{\begin{array}{l}
\mathbf{v}_\alpha(\mathbf{k}) \cdot \nabla_{\mathbf{r}} f_\alpha(\mathbf{k})+\dfrac{\mathbf{F}}{\hbar} \cdot \nabla_{\mathbf{k}} f_\alpha(\mathbf{k})=\left(\dfrac{\partial f_\alpha(\mathbf{k})}{\partial t}\right)_{\mathrm{e-ph}}+ \left(\dfrac{\partial f_\alpha(\mathbf{k})}{\partial t}\right)_{\mathrm{e-imp}}+...\\
\mathbf{v}_\lambda(\mathbf{q}) \cdot \nabla_{\mathbf{r}} n_\lambda(\mathbf{q})=\left(\dfrac{\partial n_\lambda(\mathbf{q})}{\partial t}\right)_{\mathrm{ph-ph}} + \left(\dfrac{\partial n_\lambda(\mathbf{q})}{\partial t}\right)_{\mathrm{ph-e}} + \left(\dfrac{\partial n_\lambda(\mathbf{q})}{\partial t}\right)_{\mathrm{ph-imp}} +...,
\end{array}\right. 
\end{equation}
where $\mathbf{v}_\alpha(\mathbf{k})$ and $\mathbf{v}_\lambda(\mathbf{q})$ are velocity vectors for electrons and phonons with wave vectors $\mathbf{k}$ and $\mathbf{q}$ and band index $\alpha$ and $\lambda$, respectively, $f$ and $n$ are the distribution functions for electrons and phonons and $\mathbf{F}$ is the external force, which is the electrostatic field in this work. The collision terms on the right side of the equations represent different mechanisms, including electron-phonon and electron-impurity scatterings for electrons and phonon-phonon, phonon-electron and phonon-impurity scatterings for phonons. Within a linearized BTE formalism, which only takes into account the first-order deviation of the electron and phonon distribution functions from their equilibrium values, the collision term due to electron-phonon interactions can be rewritten as~\cite{protik2022elphbolt}:
\begin{equation}\label{eqn:2}
    \left\{\begin{aligned}
\left(\dfrac{\partial f_\alpha(\mathbf{k})}{\partial t}\right)_{e-p h} \simeq &-\left[\sum_{\mathbf{k}^{\prime} \beta, \mathbf{q} \lambda} F_{\mathbf{k} \alpha}\left(\mathbf{k}^{\prime} \beta, \mathbf{q} \lambda\right)\right] \cdot \Delta f_{\mathbf{k} \alpha}+\sum_{\mathbf{k}^{\prime} \beta, \mathbf{q} \lambda}\left[F_{\mathbf{k}^{\prime} \beta}(\mathbf{k} \alpha, \mathbf{q} \lambda) \cdot \Delta f_{\mathbf{k}^{\prime} \beta}\right]+ \\ &\sum_{\mathbf{k}^{\prime} \beta, \mathbf{q} \lambda}\left[F_{\mathbf{q} \lambda}\left(\mathbf{k} \alpha, \mathbf{k}^{\prime} \beta\right) \cdot \Delta n_{\mathbf{q} \lambda}\right] \\
\left(\dfrac{\partial n_\lambda(\mathbf{q})}{\partial t}\right)_{e-p h} \simeq &\sum_{\mathbf{k} \alpha, \mathbf{k}^{\prime} \beta}\left[G_{\mathbf{k} \alpha}\left(\mathbf{k}^{\prime} \beta, \mathbf{q} \lambda\right) \cdot \Delta f_{\mathbf{k} \alpha}+G_{\mathbf{k}^{\prime} \beta}(\mathbf{k} \alpha, \mathbf{q} \lambda) \cdot \Delta f_{\mathbf{k}^{\prime} \beta}\right]-\\ &\left[\sum_{\mathbf{k} \alpha, \mathbf{k}^{\prime} \beta} G_{\mathbf{q} \lambda}\left(\mathbf{k} \alpha, \mathbf{k}^{\prime} \beta\right)\right] \cdot \Delta n_{\mathbf{q} \lambda},
\end{aligned}\right.
\end{equation}
where the coefficients $F$ and $G$ only depend on the equilibrium distribution functions of electrons $f^0$ and phonons $n^0$, 
\begin{equation} \label{eqn:3}
    \left\{\begin{array}{l}
F_{\mathbf{k} \alpha}\left(\mathbf{k}^{\prime} \beta, \mathbf{q} \lambda\right)=\left[\left(n_{\mathbf{q} \lambda}^0+f_{\mathbf{k}^{\prime} \beta}^0\right) \Pi_{-}+\left(n_{\mathbf{q} \lambda}^0+1-f_{\mathbf{k}^{\prime} \beta}^0\right) \Pi_{+}\right] \\
F_{\mathbf{k}^{\prime} \beta}(\mathbf{k} \alpha, \mathbf{q} \lambda)=\left[\left(n_{\mathbf{q} \lambda}^0+1-f_{\mathbf{k} \alpha}^0\right) \Pi_{-}+\left(n_{\mathbf{q} \lambda}^0+f_{\mathbf{k} \alpha}^0\right) \Pi_{+}\right] \\
F_{\mathbf{q} \lambda}\left(\mathbf{k} \alpha, \mathbf{k}^{\prime} \beta\right)=\left[\left(f_{\mathbf{k}^{\prime} \beta}^0-f_{\mathbf{k} \alpha}^0\right) \Pi_{-}+\left(f_{\mathbf{k}^{\prime} \beta}^0-f_{\mathbf{k} \alpha}^0\right) \Pi_{+}\right] \\
G_{\mathbf{k} \alpha}\left(\mathbf{k}^{\prime} \beta, \mathbf{q} \lambda\right)=\left[-\left(n_{\mathbf{q} \lambda}^0+f_{\mathbf{k}^{\prime} \beta}^0\right) \Pi_{-}+\left(n_{\mathbf{q} \lambda}^0+1-f_{\mathbf{k}^{\prime} \beta}^0\right) \Pi_{+}\right] \\
G_{\mathbf{k}^{\prime} \beta}(\mathbf{k} \alpha, \mathbf{q} \lambda)=\left[\left(n_{\mathbf{q} \lambda}^0+1-f_{\mathbf{k} \alpha}^0\right) \Pi_{-}-\left(n_{\mathbf{q} \lambda}^0+f_{\mathbf{k} \alpha}^0\right) \Pi_{+}\right] \\
G_{\mathbf{q} \lambda}\left(\mathbf{k} \alpha, \mathbf{k}^{\prime} \beta\right)=\left[\left(f_{\mathbf{k}^{\prime} \beta}^0-f_{\mathbf{k} \alpha}^0\right) \Pi_{-}-\left(f_{\mathbf{k}^{\prime} \beta}^0-f_{\mathbf{k} \alpha}^0\right) \Pi_{+}\right],
\end{array}\right.
\end{equation}
and
\begin{equation}
    \left\{\begin{array}{l}
\Pi_{-}=\dfrac{2 \pi}{\hbar}\left|g_{\alpha \beta \lambda}\left(\mathbf{k}, \mathbf{k}^{\prime}, \mathbf{q}\right)\right|^2 \cdot \delta\left(E_{\mathbf{k}^{\prime} \beta}-E_{\mathbf{k} \alpha}-\hbar \omega_{\mathbf{q} \lambda}\right) \cdot \delta\left(\mathbf{k}^{\prime}-\mathbf{k}-\mathbf{q}\right) \\
\Pi_{+}=\dfrac{2 \pi}{\hbar}\left|g_{\alpha \beta \lambda}\left(\mathbf{k}, \mathbf{k}^{\prime}, \mathbf{q}\right)\right|^2 \cdot \delta\left(E_{\mathbf{k}^{\prime} \beta}-E_{\mathbf{k} \alpha}+\hbar \omega_{\mathbf{q} \lambda}\right) \cdot \delta\left(\mathbf{k}^{\prime}-\mathbf{k}+\mathbf{q}\right),
\end{array}\right.
\end{equation}
where 
\begin{equation}
    g_{\alpha \beta \lambda}\left(\mathbf{k}, \mathbf{k}^{\prime}, \mathbf{q}\right)=\left(\dfrac{\hbar}{2 m_0 \omega_{\mathbf{q}_\lambda}}\right)^{1 / 2} \cdot\left\langle\mathbf{k}^{\prime} \beta\left|\partial_{\mathbf{q} \lambda} V\right| \mathbf{k} \alpha\right\rangle
\end{equation}
is the electron-phonon interaction matrix element. The non-equilibrium phonon distributions, described by $\Delta n_{\mathbf{q} \lambda}$, which appear in the electron BTE, are responsible for the phonon drag effect. When phonons are driven far away from thermal equilibrium, this term is no longer negligible and can greatly modify the electron distribution and, thus, the electronic transport properties. To capture the phonon drag effect in GaN and AlN, we utilized the Elphbolt package developed by Protik et al.~\cite{protik2022elphbolt} to solve the fully coupled electron-phonon BTEs with electron and phonon dispersions and electron-phonon coupling matrix elements all evaluated from first principles. Specifically, we used Elphbolt to transform the quantities from the real-space Wannier representation to the Bloch representation in the reciprocal space on a fine $50\times50\times50$ q-point grid and a $150\times150\times150$ k-point grid. These matrix elements were used as input to the coupled electron-phonon BTEs, where the nonequilibrium distribution functions of electrons and phonons can be solved and used to evaluate the electrical transport properties~\cite{zhou2016first}. In addition to electron-phonon scatterings, the Brooks-Herring model was employed to calculate the electron-charged-impurity scattering rates, where the impurity potential has a screened Coulomb form~\cite{brooks1955theory}.   

The solution of the distribution function of electrons in the linear response regime can be written as~\cite{zhou2016first}: 
\begin{equation}
    \begin{aligned}
        f_{\mathbf{k}\alpha} &= f^0_{\mathbf{k}\alpha} - \dfrac{\partial f^0_{\mathbf{k}\alpha}}{\partial \varepsilon_{\mathbf{k}\alpha}} ~(\mathbf{J}_{\mathbf{k}\alpha}\cdot \mathbf{E}+\mathbf{I}_{\mathbf{k}\alpha}\cdot \nabla T ) \\
        &= f^0_{\mathbf{k}\alpha} \left[1-\frac{1}{k_B T}(1-f^0_{\mathbf{k}\alpha}) (\mathbf{J}_{\mathbf{k}\alpha}\cdot \mathbf{E}+\mathbf{I}_{\mathbf{k}\alpha}\cdot \nabla T) \right] ,
    \end{aligned}
\end{equation}
where $\varepsilon_{\mathbf{k}\alpha}$ is the electron energy, $k_B$ is the Boltzmann constant, and $\mathbf{J}_{\mathbf{k}\alpha}$ and $\mathbf{I}_{\mathbf{k}\alpha}$ are the electron response coefficients of the electron state $\mathbf{k}\alpha$ to the applied electric field $\mathbf{E}$ and temperature gradient $\nabla T$, respectively.
The electrical conductivity $\sigma$ and the Seebeck coefficient $S$ can then be calculated as
\begin{equation} \begin{aligned}
    \sigma &= \dfrac{2e}{VN_kk_BT}\sum_{\mathbf{k}\alpha}\mathbf{v}_{\mathbf{k}\alpha}f^0_{\mathbf{k}\alpha}(1-f^0_{\mathbf{k}\alpha})\times \mathbf{J}_{\mathbf{k}\alpha} \\    
    \sigma S &= -\dfrac{2e}{VN_kk_BT}\sum_{\mathbf{k}\alpha}\mathbf{v}_{\mathbf{k}\alpha}f^0_{\mathbf{k}\alpha}(1-f^0_{\mathbf{k}\alpha})\times \mathbf{I}_{\mathbf{k}\alpha} ,
\end{aligned}
\end{equation}
where $V$ is the volume of the unit cell and $N_\mathbf{k}$ is the number of electronic wave vectors in the BZ.

\section{Results and Discussions}

\subsection{Phonon drag effect on carrier mobility}
\begin{figure}[!htb]
\includegraphics[scale=0.9]{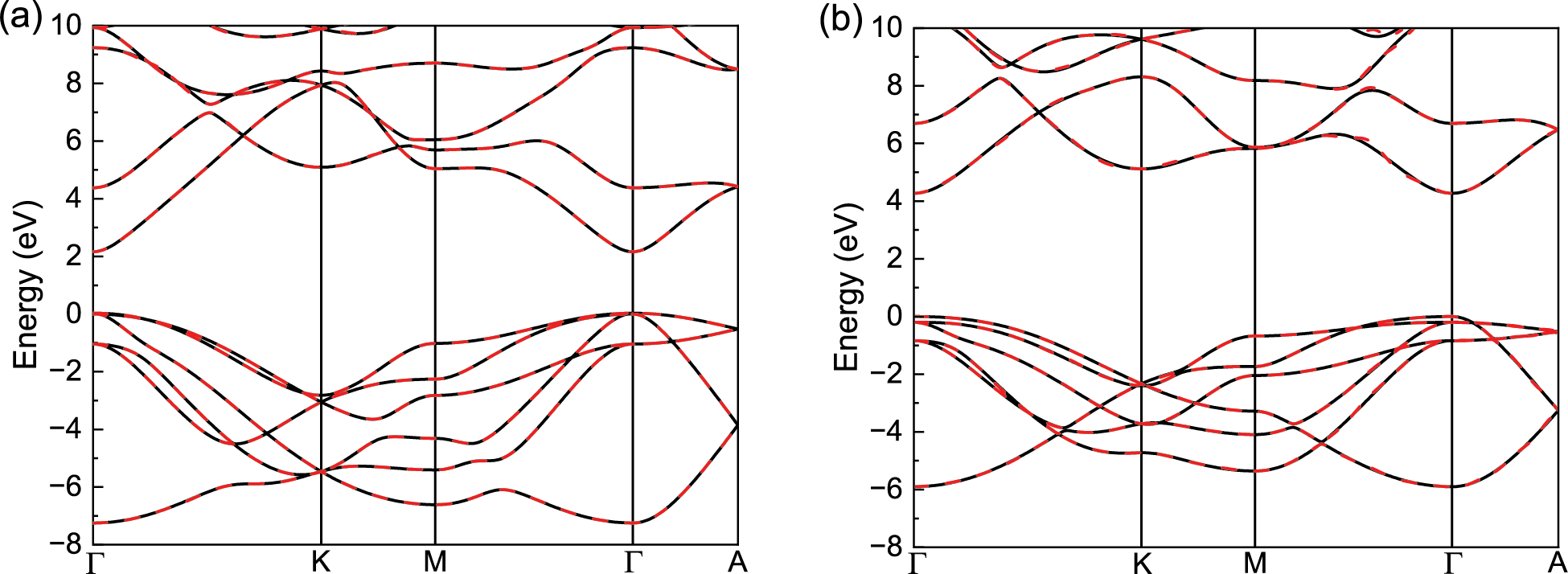}
\caption{The electronic band structure of wurtzite (a) GaN and (b) AlN. The Wannier interpolated band structures (red dashed lines) are in perfect agreement with first-principles calculations (black solid lines). } 
\label{fig:eband}
\end{figure}

The electronic band structures of wurtzite GaN and AlN were calculated using ONCV pseudopotentials within LDA, as shown in Fig.~\ref{fig:eband}. It is well established that LDA tends to underestimate the bandgap. Due to the large bandgaps in GaN and AlN, which suppresses thermal excitations and bipolar transport, accurate bandgap values are not essential in the current study. To further demonstrate the feasibility of using LDA throughout the calculation, we calculated the electron effective mass in GaN and AlN in the vicinity of the conduction-band minimum (CBM), which is transport relevant, and compared the results with the Heyd, Scuseria, and Ernzerhof (HSE) ~\cite{heyd2003hybridHSE,paier2006screened} screened hybrid functional calculations~\cite{dreyer2013effectsHSE} and quasiparticle $G_0W_0$ calculations~\cite{rinke2008consistentGW} in the literature, which were known to provide more accurate bandgap values. The results are listed in Table ~\ref{tab:table1}. Our calculation shows that the effective mass of GaN is 0.182 in the unit of free electron mass parallel to the c-axis and 0.202 perpendicular to the c-axis. In AlN, the effective mass is 0.304 along the c-axis and 0.321 perpendicular to the c-axis, which are close to the literature values. Although LDA underestimates the band gap, the similarity of electron effective mass among LDA, $G_0W_0$ and HSE justifies our usage of LDA in the calculation of electronic transport properties. The Wannier-interpolated electronic band structures are shown in Fig.~\ref{fig:eband}, which are in excellent agreement with the first-principles DFT calculation, providing a solid foundation for accurate electron-phonon matrix elements calculations. 

\newcommand{\tabincell}[2]{\begin{tabular}{@{}#1@{}}#2\end{tabular}}  
\begin{table*}
\caption{\label{tab:table1}Electron effective mass of GaN and AlN parallel and perpendicular to the c-direction in the unit of free electron mass.}
\begin{ruledtabular}
\begin{tabular}{ccccc}
 &\tabincell{c}{Direction}&\tabincell{c}{$G_0W_0$\cite{rinke2008consistentGW}}&\tabincell{c}{HSE \cite{dreyer2013effectsHSE}}&\tabincell{c}{LDA (this work) }
 \\ \hline
 GaN & $m_e^{\parallel}$&0.19&0.19& 0.182
 \\ 
 & $m_e^{\perp}$&0.21&0.22&0.202
 \\ 
 AlN & $m_e^{\parallel}$ &0.32 &0.31 &0.304
 \\ 
 & $m_e^{\perp}$&0.33&0.32&0.321
 
\end{tabular}
\end{ruledtabular}
\end{table*}

\begin{figure}[!htb]
\includegraphics[scale=0.9]{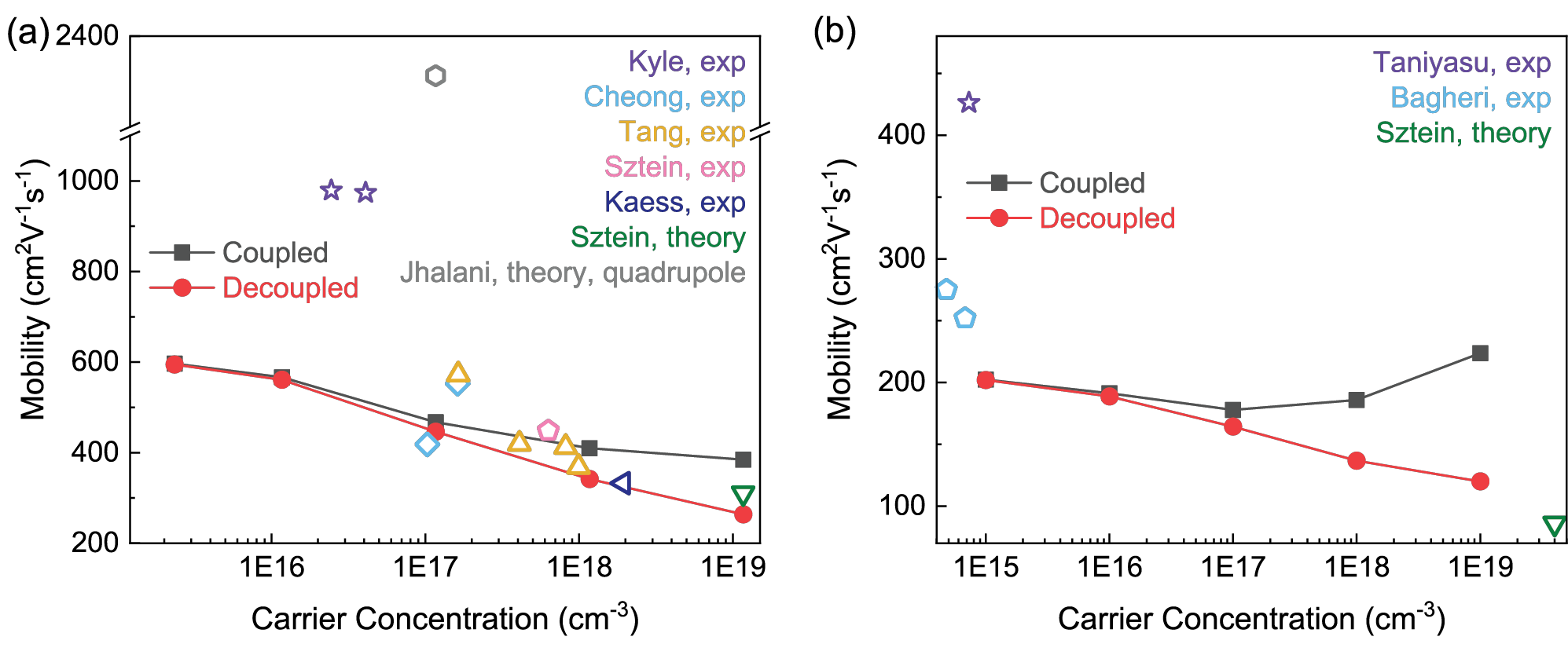}
\caption{The calculated electron mobility at room temperature with (black solid line) and without (red solid line) the phonon drag effect as a function of carrier concentration in (a) GaN and (b) AlN. The experimental and theoretical values are also shown for comparison. Experimental mobility values of GaN are taken from Ref.~\cite{cheong2000conductiveGaNm1,kyle2014highGaNm2,tang1999growthGaNm3,sztein2014polarizationGaNm4,kaess2016correlationGaNm5}. The experimental values of AlN are taken from Ref.~\cite{bagheri2022highAlN300,taniyasu2006increasedAlN426}. The theoretical BTE result is taken from Ref.~\cite{sztein2013calculatedAlNBTE}, which does not consider the phonon drag effect. The first-principles calculation of GaN including the quadrupolar term is taken from Ref.~\cite{jhalani2020piezoelectricQD}. }
\label{fig:Mobility}
\end{figure}

First, we focus on the enhancement of carrier mobility in GaN and AlN due to the phonon drag effect. When the phonon drag effect is not considered, the electron-phonon interaction is a purely momentum-dissipation process for electrons that limits the electron mobility. Microscopically, however, the electron-phonon interaction process conserves the total momentum and electrons transfer their momentum to phonons, creating a nonequilibrium phonon distribution. While a fraction of the excess momentum that phonons receive from electrons will be dissipated through anharmonic phonon-phonon interactions and phonon-impurity scatterings, the rest can be pumped back into electrons through electron-phonon interactions, which can act to boost the carrier mobility. This effect has been observed in previous phonon drag studies in GaAs~\cite{protik2020coupledGaAs} and 3C-SiC~\cite{protik2020electronSiC}. Here, we evaluated this effect on the carrier mobility in GaN and AlN, which is of paramount importance for device applications. The room-temperature electron mobility of n-type GaN and n-type AlN were calculated with the electron concentrations ranging from $\rm 10^{15}~cm^{-3}$ to $\rm 10^{19}~cm^{-3}$. 
In Fig.~\ref{fig:Mobility}, we show the electron mobility as a function of electron concentration at room temperature with and without the phonon drag contribution. The experimental values~\cite{cheong2000conductiveGaNm1,kyle2014highGaNm2,tang1999growthGaNm3,sztein2014polarizationGaNm4,kaess2016correlationGaNm5,bagheri2022highAlN300,taniyasu2006increasedAlN426} and the theoretical BTE result without the phonon drag contribution~\cite{sztein2013calculatedAlNBTE} are also plotted here for comparison. It is noted that the experimental values of the electron mobility in GaN at $n=10^{16}~\rm cm^{-3}$ are around 1000\,$\rm cm^2/V\,s$~\cite{kyle2014highGaNm2}, while our calculation predicts 600\,$\rm cm^2/V\,s$. This discrepancy is due to the fact that our current electron-phonon interaction calculations only included the dipole-like Fr\"{o}lich long-range coupling~\cite{frohlich1954electronsFRO}, where the higher-order quadrupolar term was excluded. Jhalani et al. has shown that considering only the dipole interaction overestimates the electron interactions with acoustic phonons in polar, and particularly, piezoelectric materials like GaN and AlN, and the inclusion of the quadrupole term can correct this overestimation and provide more accurate coupling matrix elements between electrons and acoustic phonons~\cite{jhalani2020piezoelectricQD}. Their result is also labeled in Fig.~\ref{fig:Mobility}(a) for comparison. Although our calculations underestimated the electron mobility of GaN at lower carrier concentrations, the results at higher carrier concentrations were in better agreement with the experiments, since the quadrupole term primarily impacts the interactions between electrons and acoustic phonons with small wave vectors, which represents a smaller fraction of the total electron-phonon scatterings at higher concentrations. It is shown here that, without taking the phonon drag effect into consideration, the calculated mobility (labeled ``decoupled'' in Fig.~\ref{fig:Mobility}) would be lower than the experimental values, which are usually further limited by the sample quality and other nonidealities, implying that the phonon drag is crucial for accurate calculation of the mobility at higher electron concentrations. Although the experimental values of mobility in AlN are scarce in the literature given the difficulty in making highly n-doped AlN samples, our results are in reasonable agreement with the reference values that can be found. 

\begin{figure}[!htb]
\includegraphics[scale=0.4]{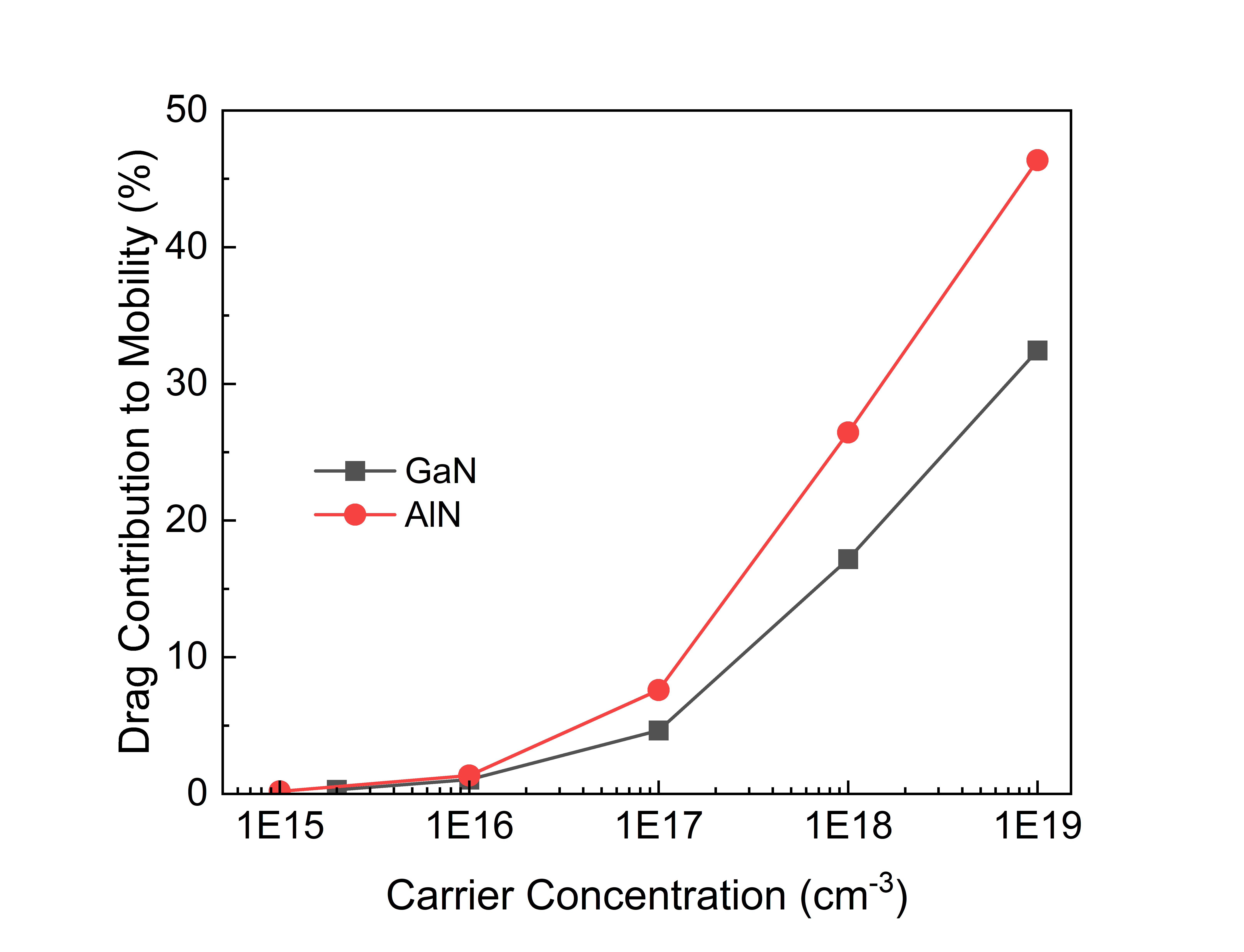}
\caption{The phonon drag contribution to the total electron mobility in GaN and AlN. The phonon drag contribution becomes more prominent with the increasing carrier concentration.} 
\label{fig:dragM}
\end{figure}

\begin{figure}[!htb]
\includegraphics[scale=0.9]{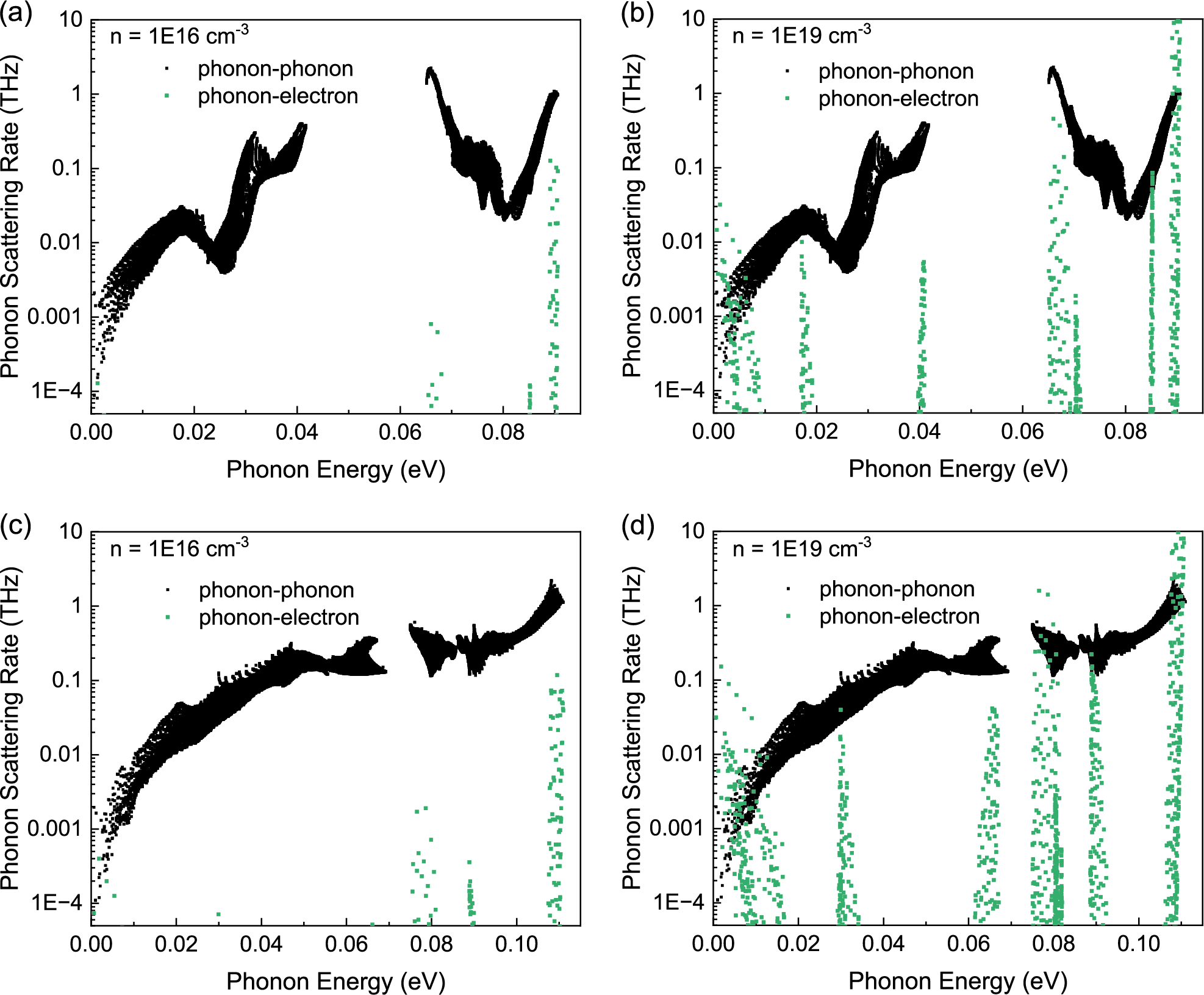}
\caption{Calculated phonon-phonon and phonon-electron scattering rates of (a) GaN at $n=10^{16}~\rm cm^{-3}$; (b) GaN at $n=10^{19}~\rm cm^{-3}$; (c) AlN at $n=10^{16}~\rm cm^{-3}$ and (d) AlN at $n=10^{19}~\rm cm^{-3}$. The strong phonon-electron scattering of the low-frequency acoustic phonons and longitudinal optical phonons at higher concentrations facilitates the momentum circulation between electrons and phonons.  } 
\label{fig:scattering}
\end{figure}

The percentage of the phonon drag contribution to the electron mobility in GaN and AlN is shown in Fig.~\ref{fig:dragM}. For both GaN and AlN, the phonon drag contribution is negligible at $n=10^{15}\rm ~cm^{-3}$. However, the phonon drag contribution becomes more prominent with an increasing carrier concentration. At $n=10^{19}~\rm cm^{-3}$, 32.4\% of the total electron mobility is due to the phonon drag in GaN, and 46.4\% in AlN. This carrier concentration dependence can be explained by the relative strength between phonon-phonon scatterings and phonon-electron scatterings. The phonon scattering rates of GaN and AlN within RTA at $n=10^{16}~\rm cm^{-3}$ and $n=10^{19}~\rm cm^{-3}$ are shown in Fig.~\ref{fig:scattering}, where the black dots denote the phonon-phonon scattering rates and the green dots denote the  phonon-electron scattering rates. Although the phonon drag effect is not included at the RTA level, the phonon scattering rates within RTA can provide useful information for our analysis. It is seen that at low carrier concentrations, the phonon-electron scattering is much weaker than the phonon-phonon scattering, while at high carrier concentrations, the phonon-electron scattering becomes stronger than the phonon-phonon scattering for the the low-frequency acoustic phonons and the polar longitudinal optical (LO) phonons, since the number of available electronic states that phonons can couple with increases~\cite{liao2015significant,quan2021impact}. This strong phonon-electron scattering facilitates the momentum circulation between electrons and phonons. The momentum previously transferred to phonons from electrons can be pumped back more effectively at higher carrier concentrations, making electrons less dissipative compared with the case without the phonon drag effect. This suggests that, as the carrier concentration increases, although the overall carrier mobility decreases due to the increased electron-phonon scattering, the relative enhancement from the phonon drag effect increases due to more effective electron-phonon momentum circulation.

\begin{figure}[!htb]
\includegraphics[scale=0.4]{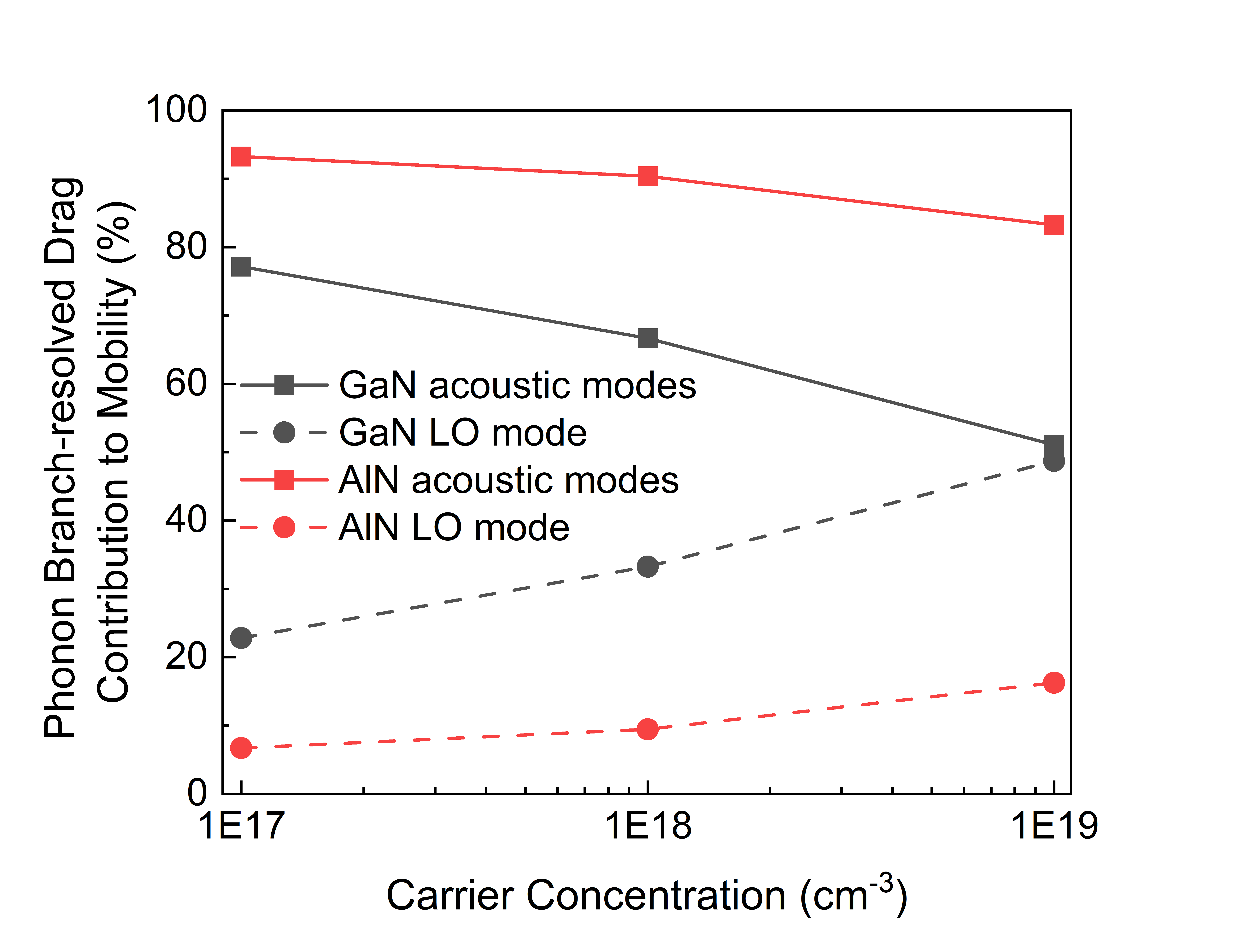}
\caption{Phonon-branch-resolved contribution to the drag-enhanced electron mobility in GaN and AlN. Black lines denote the results in GaN and red lines denote the results in AlN. Solid lines represent the acoustic modes contribution and the dashed lines represent the LO mode contribution. The contributions from other optical phonon modes are negligible and not plotted here.} 
\label{fig:mode}
\end{figure}

The details of the phonon drag influence on mobility can be further illustrated by analyzing the percentage contribution of each phonon branch to the enhanced mobility due to the drag effect, as shown in Fig.~\ref{fig:mode}. The acoustic modes' contribution is depicted by the solid lines, while the LO phonon mode contribution is represented by the dashed lines. The contributions from other optical modes are negligibly small and, thus, are not shown here. The strong interaction between LO phonons and electrons in both GaN and AlN results in a significant contribution to the drag-induced mobility enhancement, making the LO phonons' impact on mobility non-negligible. Specifically, at a carrier concentration of $n=10^{19}~\rm cm^{-3}$, the contribution of LO phonons to the mobility gain is almost equal to that of the acoustic phonons. As discussed previously, the exclusion of quadrupolar correction in our calculation tends to overestimate the electron interaction with acoustic phonons~\cite{jhalani2020piezoelectricQD}, leading to inaccuracies in the mobility calculation. However, since the LO phonon scattering with electrons is largely controlled by the Fr\"{o}hlich dipolar interaction rather than the quadrupolar term~\cite{jhalani2020piezoelectricQD}, this significant drag effect from LO phonons is accurately captured in our calculation, signaling the important role of polar LO phonons in the phonon drag effect in strongly polar semiconductors. This observation is also a manifestation of the fact that the impact of the phonon drag on the carrier mobility mainly depends on the electron-phonon scattering rate. This is in contrast to the phonon drag contribution to the Seebeck coefficient, which also depends on the phonon mean free paths. This distinction will be discussed in more detail in the next section. 

\subsection{Phonon drag effect on Seebeck coefficient}
\begin{figure}[!htb]
\includegraphics[scale=0.9]{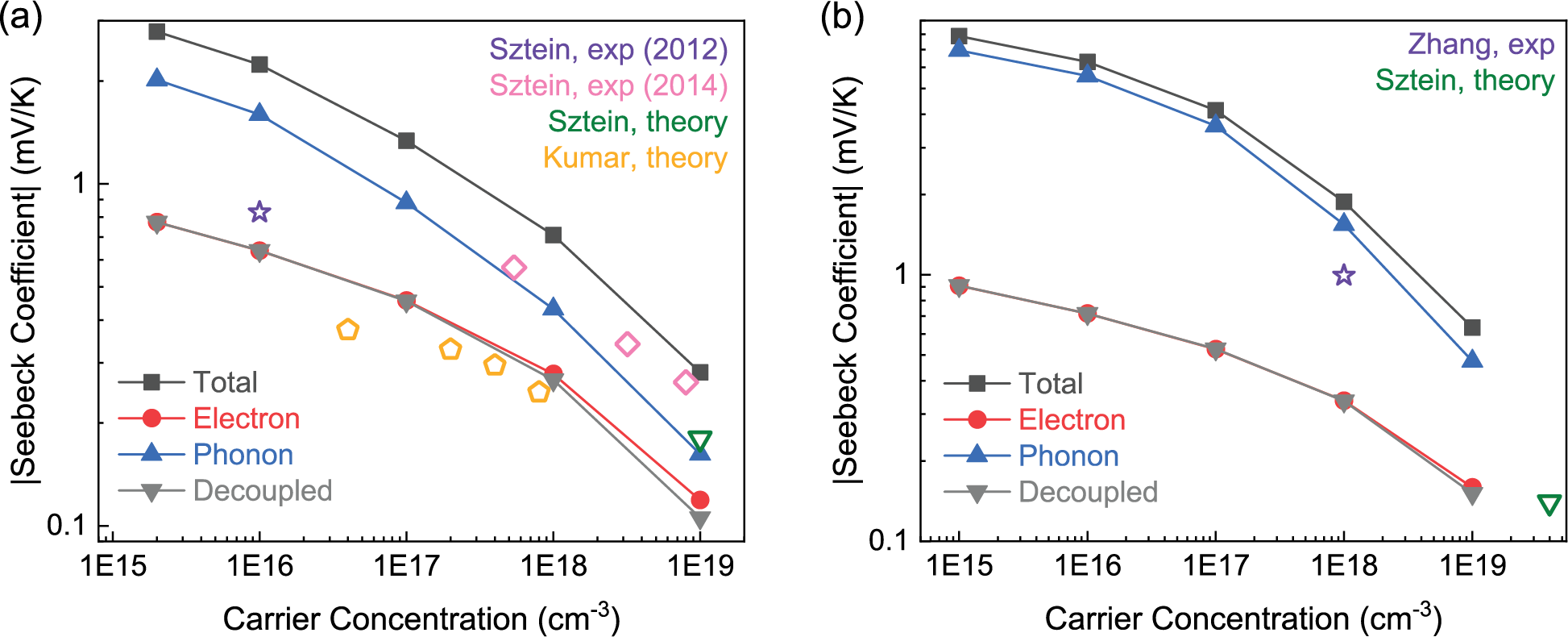}
\caption{The calculated Seebeck coefficient with and without considering the phonon drag as a function of carrier concentration in (a) GaN and (b) AlN. The experimental values of the Seebeck coefficient in GaN are taken from Ref.~\cite{sztein2014polarizationGaNm4,sztein2012thermoelectricGaNS2} and the theoretical calculation results are from Ref.~\cite{sztein2013calculatedAlNBTE,kumar2021thermoelectricGaNS4}. The experimental values of the Seebeck coefficient in AlN is taken from Ref.~\cite{zhang2011highAlNs} and the theoretical BTE calculation result is from Ref.~\cite{sztein2013calculatedAlNBTE}. The calculation results cited here did not take the phonon drag effect into consideration.} 
\label{fig:Seebeck}
\end{figure}

The most prominent manifestation of the phonon drag effect is its impact on the Seebeck coefficient. Here, the Seebeck coefficient of n-doped GaN and AlN at room temperature was calculated with and without considering the phonon drag contribution. The absolute value of the Seebeck coefficient as a function of carrier concentration is shown in Fig.~\ref{fig:Seebeck}. The experimental values~\cite{sztein2014polarizationGaNm4,sztein2011highGaNS1,sztein2012thermoelectricGaNS2,zhang2011highAlNs} and previous theoretical calculation results~\cite{kumar2021thermoelectricGaNS4,sztein2013calculatedAlNBTE} are also shown for comparison. Our results from solving the decoupled electron and phonon BTEs, shown as the gray line, are in agreement with previous theoretical calculations using analytical models without considering the phonon drag effect. It is noted that due to the exclusion of the quadrupolar corrections~\cite{jhalani2020piezoelectricQD}, our calculated phonon drag contribution can be overestimated at low carrier concentrations because of the overestimated electron-phonon matrix elements involving low-frequency acoustic phonons. At higher carrier concentrations, where a broader spectrum of phonons contribute to electron-phonon interactions and the overestimated part accounts for less of the total electron-phonon scatterings, our results are in good agreement with the experimental results. Since the calculated Seebeck coefficient without considering the phonon drag is much lower than the experiments, in which samples are also affected by high threading dislocation densities and other impurities, the phonon drag effect is indispensable for accurately predicting the Seebeck coefficient in GaN and AlN. We also found that, different from the electron mobility, to which the phonon drag contribution is negligible at low carrier concentrations, the influence of the phonon drag on the Seebeck coefficient is prominent throughout the entire carrier concentration range that we investigated. It should also be mentioned that, at high carrier concentrations, in addition to electron-impurity scattering, phonon-impurity scattering also needs to be considered. The typical method for n-type doping in GaN and AlN is adding silicon, which acts as a shallow donor~\cite{breckenridge2021highSi-AlN,boguslawski1997dopingSi-GaN}, meaning that the carrier concentration is quite close to the concentration of silicon atoms at room temperature. At $n=10^{19}~\rm cm^{-3}$, our calculation found that the inclusion of the silicon atoms with the same concentration only leads to less than 1\% difference compared with the result without including the phonon-impurity scattering.

\begin{figure}[!htb]
\includegraphics[scale=0.4]{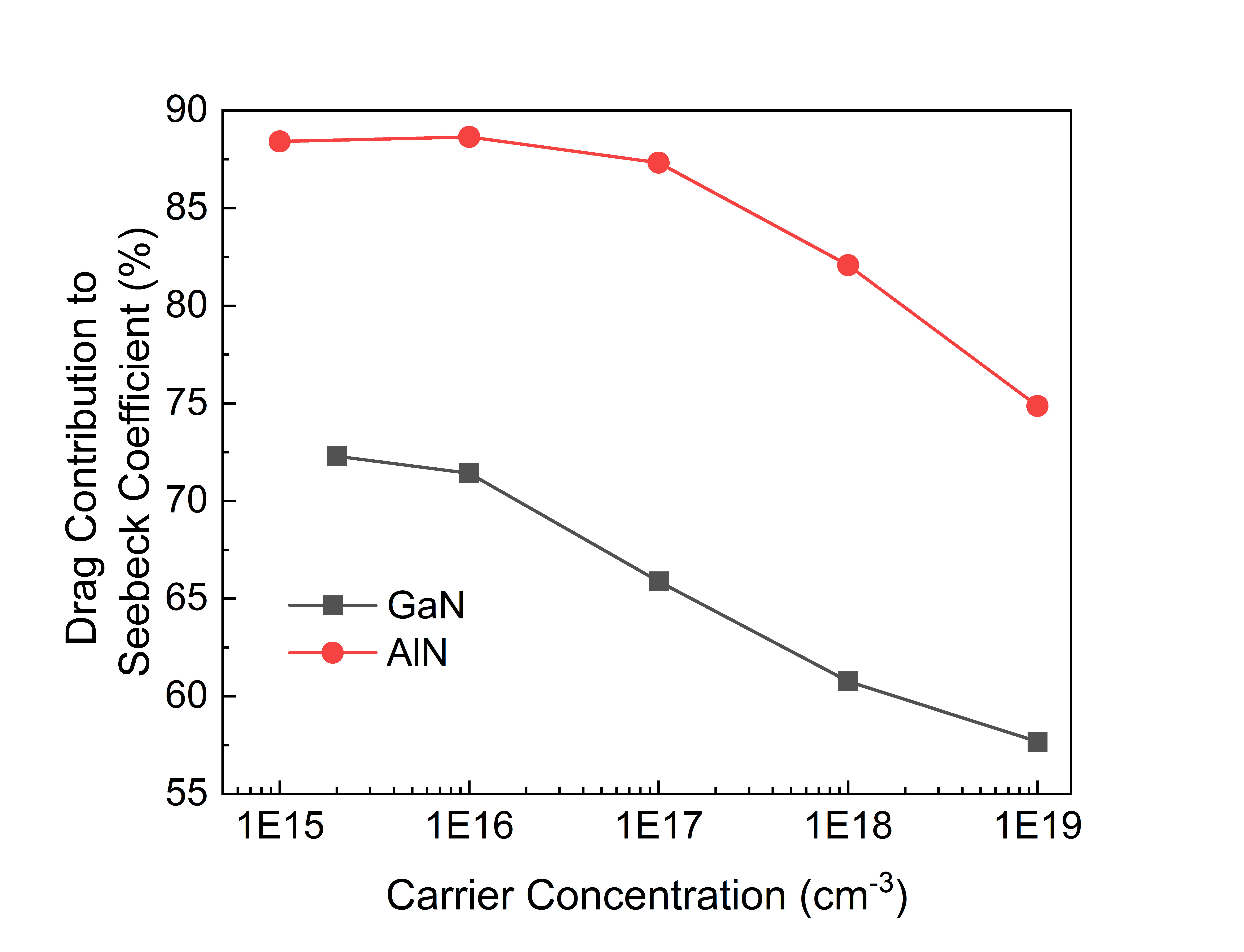}
\caption{The calculated phonon drag contribution to the total Seebeck coefficient as a functino of the carrier concentration in GaN and AlN. As the carrier concentration increases, the phonon drag component changes little at first and then decreases rapidly. Despite this trend, the phonon drag impact remains significant across the entire range of carrier concentrations.} 
\label{fig:dragS}
\end{figure}

To gain more insight into the phonon drag contribution to the Seebeck coefficient, the percentage phonon drag contribution is plotted as a function of the carrier concentration, as shown in Fig.~\ref{fig:dragS}. It is found that at relatively low carrier concentrations, the phonon drag contribution changes very little with the carrier concentration, whereas at higher concentrations, the drag contribution decreases rapidly with an increasing carrier concentration. This trend is also observed and discussed in silicon~\cite{zhou2015ab} and SiC~\cite{protik2020electronSiC}, where the phonon drag contribution to the Seebeck coefficient is expected to be constant at low carrier concentrations. According to Herring's theory~\cite{herring1954theory}, the drag contribution to the Seebeck coefficient is proportional to the momentum gain per electron from nonequilibrium phonons. At low carrier concentrations, it is the number of available electronic states to participate in electron-phonon interactions that limits the momentum transfer from phonons to electrons. In such a case, the momentum gain per electron does not change significantly with the carrier concentration. In contrast, at higher carrier concentrations, the number of electronic states that are able to couple with phonons is no longer a limiting factor. Instead, with the increased phonon-electron scattering rate, the shortened phonon relaxation time becomes the limiting factor, since phonons with shorter lifetimes imply more likelihood to be scattered before the momentum can be transferred to the electron system. As a result, the total electron momentum gain saturates, and because of the increased carrier concentration, the momentum gain per electron decreases, leading to the rapid reduction of the phonon drag contribution at higher concentrations. The above interpretation is based on the Seebeck picture, in which phonon drag facilitates more electrons traveling along the temperature gradient by transferring momentum from non-equilibrium phonons to electrons. In parallel, the equivalent Peltier picture, in which an isothermal electric field induces a heat flow, can provide a more straightforward interpretation regarding the phonon and the electron contribution to the Seebeck coefficient. Since phonons cannot directly couple to the electric field, the nonzero Seebeck contribution from phonons is completely a result of the drag effect. Given the Kelvin relation\cite{sondheimer1956kelvin}, an extra contribution to the Peltier coefficient implies the same extra contribution to the Seebeck coefficient. In the Peltier picture, the electric field generates a non-equilibrium distribution of electrons, which transfer momentum to phonons through electron-phonon interactions so that phonons can also contribute to the total Seebeck coefficient. At low carrier concentrations, the phonon contribution is almost independent on the carrier concentration, since the phonon lifetimes remain the same due to the dominant phonon-phonon scatterings. At higher carrier concentrations, phonon lifetimes decrease due to stronger phonon-electron scatterings and more momentum is returned back to electrons, leading to the decrease of the phonon contribution to the Seebeck coefficient. Although the phonon drag percentage contribution decreases with an increasing carrier concentration, as shown in Fig.~\ref{fig:dragS}, the overall contribution is still significant throughout the carrier concentration range, stemming from the weak anharmonic phonon-phonon scattering and, thus, a low dissipation rate of excess phonon momentum in GaN and AlN. 

\begin{figure}[!htb]
\includegraphics[scale=0.9]{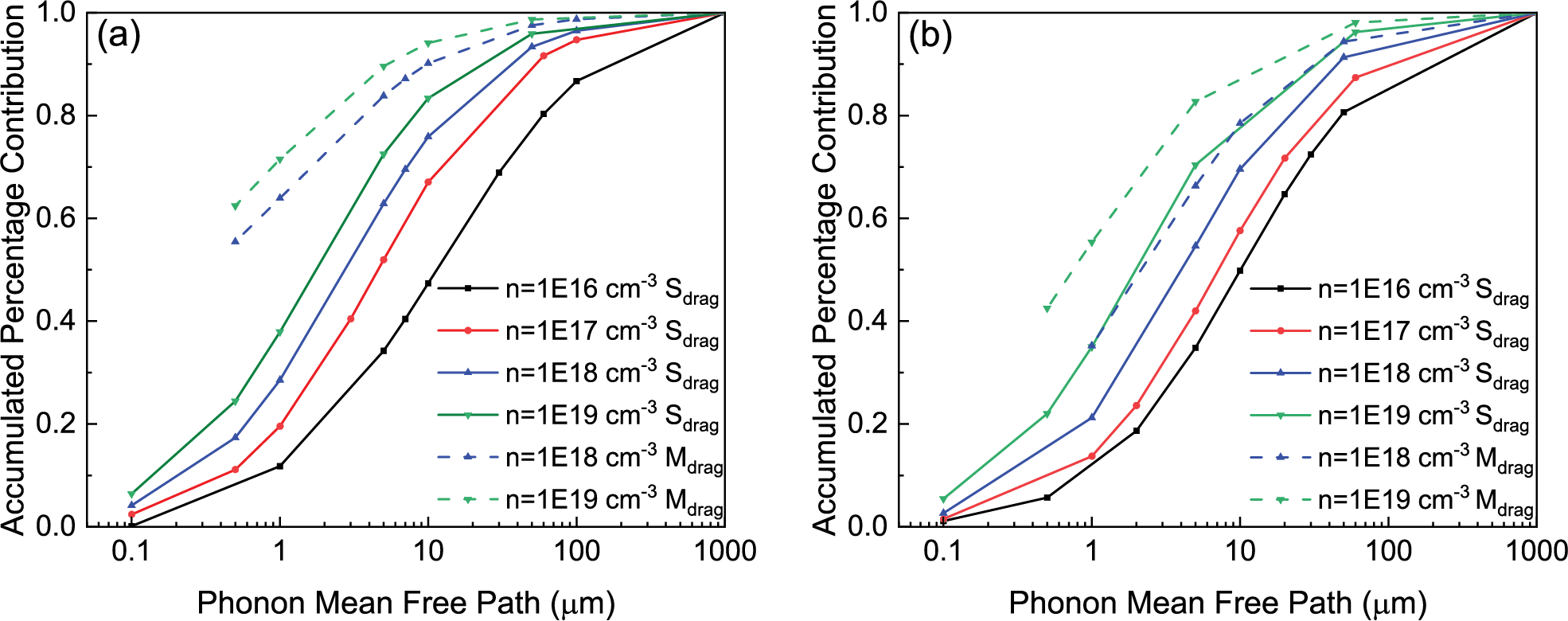}
\caption{The calculated accumulated percentage contribution of the phonon drag effect to the Seebeck coefficient (solid lines) and the mobility (dashed lines) as a function of the phonon mean free paths in (a) GaN and (b) AlN. Phonons with shorter mean free paths tend to contribute more to the mobility enhancement than to the Seebeck coefficient.} 
\label{fig:MFP}
\end{figure}

\begin{figure}[!htb]
\includegraphics[scale=0.9]{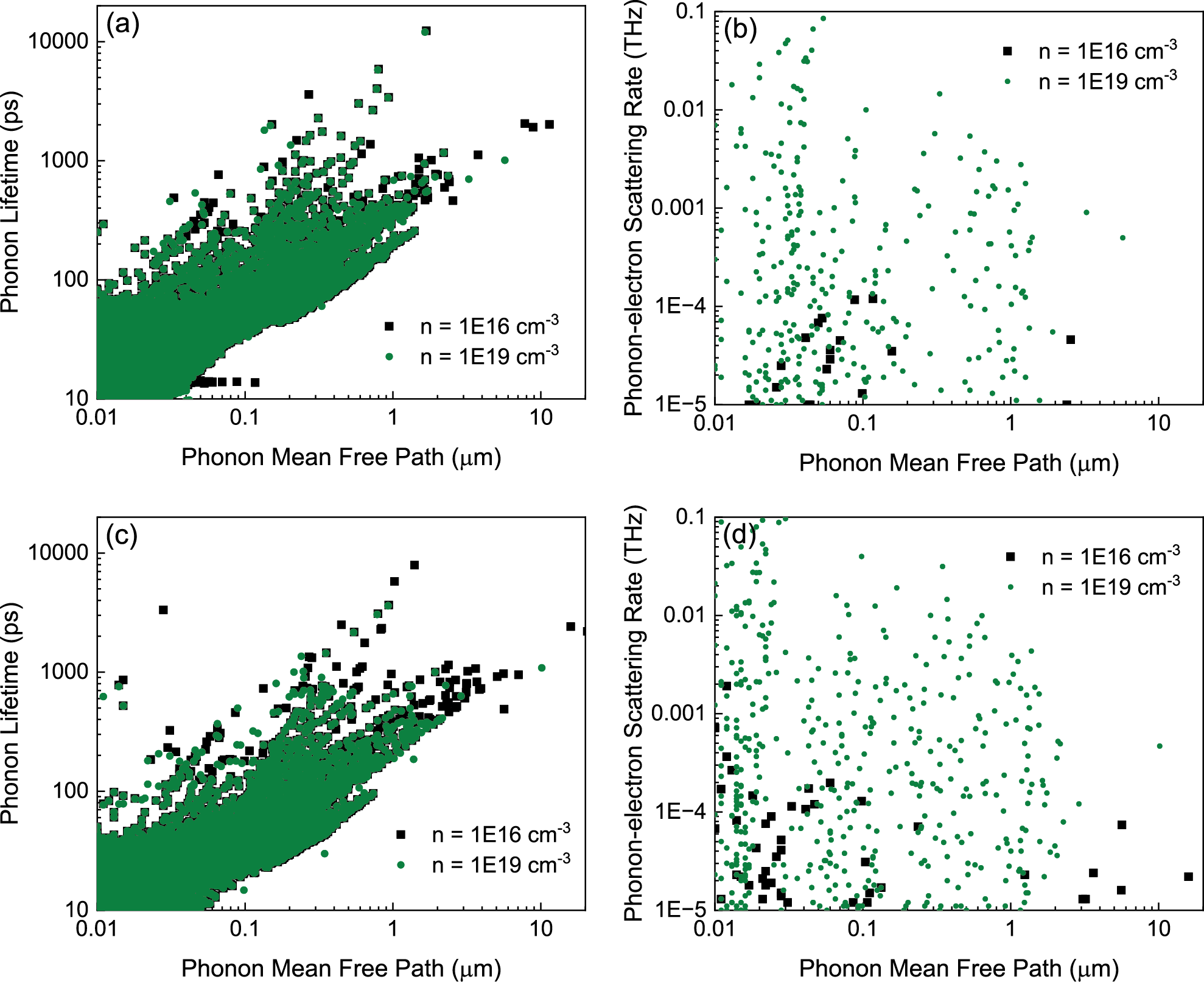}
\caption{The calculated phonon lifetime (left column) and phonon-electron scattering rate (right column) as a function of the phonon mean free path in GaN (top panels) and AlN (bottom panels).  } 
\label{fig:sc-MFP}
\end{figure}

To further understand the roles played by different phonon modes in the phonon drag effect, we also calculated the accumulated contribution to the phonon drag enhancement of the Seebeck coefficient and the carrier mobility with respect to the phonon mean free path (MFP), as shown in Fig.~\ref{fig:MFP}. This calculation was conducted by excluding the contribution from phonons with MFPs beyond a given threshold, which is labeled in the horizontal axis in Fig.~\ref{fig:MFP}. The mobility accumulation curves at lower carrier concentrations are not shown here, since the mobility gain due to phonon drag is quite small at lower carrier concentrations. It is found that at the same carrier concentration and the same phonon MFP threshold, the percentage contribution of the phonon drag effect to the mobility enhancement is greater than that to the Seebeck coefficient enhancement, indicating that phonons with shorter MFPs contribute more to the carrier mobility than to the Seebeck coefficient. This distinction arises again from the different mechanisms of the phonon drag influence on mobility and Seebeck coefficient. The increase in mobility due to phonon drag originates from the extra momentum that electrons acquire through interactions with nonequilibrium phonons. With stronger electron-phonon interactions, more momentum is transferred back to electrons and, thus, more mobility gain can be achieved. In other words, the mobility gain due to phonon drag is only determined by the electron-phonon interaction strength, and the lifetime of drag-active phonons is not a contributing factor. In contrast, a large phonon drag contribution to the Seebeck coefficient requires not only strong electron-phonon interactions but also weak phonon momentum dissipation due to phonon-phonon and phonon-electron scatterings, as discussed in the Peltier picture above. For instance, as can be seen in Fig.~\ref{fig:MFP}, at $n= 10^{19}~\rm cm^{-3}$, phonons with MFP shorter than $0.5\,\mu$m, whose lifetime is shorter than 1000\,ps as shown in Fig.~\ref{fig:sc-MFP}(a) and (c), contribute to more than 60\% and 40\% of the total mobility gain in GaN and AlN, whereas these short-lived phonons only contribute to 24\% and 22\% of the total Seebeck gain in GaN and AlN, respectively. We also found that the accumulated phonon drag contribution to the Seebeck coefficient and the mobility shifts toward the shorter phonon MFP region as the carrier concentration increases. This MFP-related feature is due to momentum and energy conservation in electron-phonon interactions. At low carrier concentrations, only the phonons with small wave vectors can couple with electronic states close to the band edges while satisfying the energy conservation condition. Typically, phonons with smaller wave vectors possess longer MFPs. With the increase of the carrier concentration, electrons occupy more of the reciprocal space and the energy and momentum conservation requirements in electron-phonon interactions become easier to satisfy, making phonons with shorter MFP active contributors to the phonon drag, as shown in Fig.~\ref{fig:sc-MFP}(b) and Fig.~\ref{fig:sc-MFP}(d).

\subsection{Phonon drag effect in p-type GaN and AlN}
For the completeness of the study, we also calculated the phonon drag contribution in p-type GaN and AlN in both low and high doping regimes, despite the fact that effective p-type doping remains challenging experimentally in both materials~\cite{sarkar2018n}. The results are summarized in Table~\ref{tab:table2}. The heavy effective mass of the valence band results in a greater number of electronic states available for scattering, which makes the calculation computationally intensive and time-consuming. In our calculation, a $30\times30\times30$ q-point grid and a $90\times90\times90$ k-point grid were used to transform the electron-phonon scattering matrix elements from the real-space Wannier representation to the Bloch representation in the reciprocal space. The convergence of our calculation with respect to the sampling mesh density was checked. Similar to the results in n-type materials, the contribution of the phonon drag effect to mobility is negligible at the low carrier concentration, while it has a great impact on the Seebeck coefficient. However, at the high carrier concentration, the phonon drag effect leads to a noticeable increase in the mobility and also has a significant contribution to the Seebeck coefficient enhancement. 

\begin{table*}
\caption{\label{tab:table2}Electron mobility and the Seebeck coefficient of p-type GaN and AlN.}
\begin{ruledtabular}
\begin{tabular}{cccc}
 &\tabincell{c}{Carrier\\ Concentration \\($\rm cm^{-3}$)}&\tabincell{c}{Drag Contribution \\to Mobility (\%)}&\tabincell{c}{Drag Contribution to \\ Seebeck Coefficient (\%)}
 \\ \hline
 GaN & $\rm 10^{16}$&0.3&90.1
 \\ 
 & $\rm 10^{19}$&25.7&82.7
 \\ 
 AlN & $\rm 10^{16}$ &1.0&94.0
 \\ 
 & $\rm 10^{19}$&27.3&80.1
 
\end{tabular}
\end{ruledtabular}
\end{table*}

\section{Conclusion}
In summary, we have studied the phonon drag effect on electrical transport properties in wide bandgap GaN and AlN by solving the fully coupled electron-phonon BTEs. The electron mobility and the Seebeck coefficient were calculated as a function of carrier concentration ranging from $\rm 10^{15}~cm^{-3}$ to $\rm 10^{19}~cm^{-3}$. We found that, even at room temperature, the phonon drag effect is prominent in both the mobility and the Seebeck coefficient. Significant enhancements of both the carrier mobility and the Seebeck coefficient were observed especially at high carrier concentrations. The strong electron-phonon coupling strength can enhance momentum transfer between electrons and phonons, and the weak anharmonic phonon-phonon scattering is beneficial for the significant enhancement of the Seebeck coefficient. Our findings highlight the importance of including the phonon drag effect in accurately predicting the electrical transport properties not only in wide bandgap group-III nitrides but also in strongly polar semiconductors in general.

\begin{acknowledgments}
We are grateful to Dr. Nakib H. Protik for assistance with the Elphbolt code. This work is based on research supported by the U.S. Air Force Office of Scientific Research under award number FA9550-22-1-0468 and the National Science Foundation (NSF) under award number CBET-1846927. Y.Q. and Y.C. also acknowledge the support from the Graduate Traineeship Program of the NSF Quantum Foundry via the Q-AMASE-i program under award number DMR-1906325 at the University of California, Santa Barbara (UCSB). This work used Stampede2 at Texas Advanced Computing Center (TACC) and Expanse at San Diego Supercomputer Center (SDSC) through allocation MAT200011 from the Advanced Cyberinfrastructure Coordination Ecosystem: Services \& Support (ACCESS) program, which is supported by National Science Foundation grants 2138259, 2138286, 2138307, 2137603, and 2138296. Use was also made of computational facilities purchased with funds from the National Science Foundation (award number CNS-1725797) and administered by the Center for Scientific Computing (CSC) at UCSB. The CSC is supported by the California NanoSystems Institute and the Materials Research Science and Engineering Center (MRSEC; NSF DMR-1720256) at UCSB.
\end{acknowledgments}

\bibliography{references.bib}
\clearpage

\end{document}